# Why Condorcet Consistency is Essential

**Richard B. Darlington**
**Cornell University**

## Abstract

In a single-winner election with several candidates and ranked-choice or rating-scale ballots, a Condorcet winner is one who wins all their two-way races by majority rule (MR). A voting system has Condorcet consistency (CC) if it names any Condorcet winner the winner. Many voting systems lack CC, but a three-step line of reasoning is used here to show why it's necessary. In step 1 we show that we can dismiss all the electoral criteria which conflict with CC. In step 2 we point out that CC follows almost automatically if we can agree that MR is the only acceptable system for two-candidate elections. In step 3 we make that argument for MR. This argument itself has three parts: (1) In two-candidate races, the only well-known alternatives to MR can sometimes name as winner a candidate who is preferred over their opponent by only one voter, with all others preferring the opponent. That's unacceptable. (2) Those same systems are also extremely susceptible to strategic insincere voting. (3) In simulation studies using spatial models with two candidates, the best-known alternative to MR picks the best (most centrist) candidate significantly less often than MR does.

## Overview

When rating scales or ranked-choice ballots are used to elect a single winner from several candidates, a Condorcet winner is one who wins all their two-way races by majority rule. A voting system (a rule for picking a winner) is said to have Condorcet consistency (CC) if it names any Condorcet winner as the winner. Many electoral theorists consider Condorcet consistency essential. But in Felsenthal's (2012) review of 18 prominent voting systems for multi-candidate elections, only 8 have CC. And when Laslier (2012) asked 22 prominent electoral theorists which voting systems they would recommend, the two systems with the most recommendations were approval voting and the Hare system, neither of which has CC.

This paper defends the necessity of CC in two different ways. First it attacks five electoral criteria which seem plausible on the surface but conflict with CC. By showing that these criteria can all be dismissed, it clears the way for accepting CC. Second, this paper shows why majority rule (MR) is the only acceptable voting system for two-candidate elections. That view is not universally accepted, as we shall see. But if MR is so accepted, then the need for CC follows almost automatically, because a Condorcet winner wins all their two-way races by MR. So we defend MR as the only acceptable two-candidate system.

The minimax voting system is a well-known Condorcet-consistent system; this paper mentions it several times. Minimax names any Condorcet winner as a winner. If there is no such winner, each candidate's largest loss (LL) in two way-races is found, and the candidate with the smallest LL is named the winner.

## Dismissing five electoral criteria which conflict with CC

Table 3.2 of Felsenthal (2012, p. 33) names five electoral criteria which are violated by all eight of the Condorcet-consistent systems which Felsenthal considered important enough to mention: minimax, Dodgson, Black, Copeland, Kemeny, Nanson, Schwartz, and Young. Thus for all practical purposes, all five of these criteria conflict with Condorcet consistency. The Felsenthal table calls these five criteria no-show; twin; truncation; reinforcement or inconsistency or multiple districts; and violation of the subset choice condition (SCC). A voting system violates the no-show criterion if a voter can sometimes benefit by abstaining rather than voting sincerely. The twin criterion is similar to no-show; it prohibits an anomaly in which two people who rank candidates the same find it beneficial for one of them to vote and the other to abstain. The truncation criterion is also somewhat similar; it is violated if a voter can benefit by listing only their first few choices rather than all of them. The multiple-districts criterion is violated if a candidate could win in each of two districts but lose if the two districts are merged into one. SCC is violated if one candidate could win, but lose if one of the losing candidates dropped out and votes were recounted with that candidate omitted.

I'll call a criterion an *optimizing* criterion if its main purpose seems to be to promote selection of the candidates who best reflect the popular will. I'll call a criterion an *anti-manipulation* criterion if a major purpose of the criterion seems to be to reject voting systems which allow some sort of electoral manipulation. The no-show, twin, and truncation criteria meet the latter definition. Many anti-manipulation criteria can also be considered to be optimizing. For instance, it may be hard to believe that a system failing any of these three criteria would generally select the best candidates even in the absence of any attempts at manipulation, so those criteria are also optimizing criteria. It's nevertheless useful to distinguish between the two types of criterion, since some criteria are solely optimizing criteria and others are not.

The anti-manipulation roles of these three criteria seem unimportant when they are applied to Condorcet-consistent systems, because these systems violate those criteria only if there is a Condorcet paradox and the would-be manipulators know how others will vote or have voted. Those two conditions together make the problem so rare as to be inconsequential. We should still address these criteria as optimizing criteria. But the whole purpose of optimizing criteria is to promote selection of the best candidates. Thus when one optimizing criterion conflicts with another (and these all conflict with Condorcet consistency), it seems reasonable to use computer simulations to see which criteria lead to selection of the best candidates. Section 5 of Darlington (2016) describes a whole series of simulation studies. All found that the best candidates were selected by minimax, a voting system which violates all five criteria of this section because the system is Condorcet-consistent. That suggests dismissing all five of these criteria. But there are additional reasons for dismissing SCC and the multiple-districts criterion.

SCC is sometimes called IIA for "independence of irrelevant alternatives." That term helps us see what's wrong with SCC/IIA, because it confuses the "irrelevant alternative" (the candidate who withdrew) with the data collected because that candidate was in the race. To see why, imagine a league with three teams in which each pair of teams plays 9 games. Team A has won all of its games against B, and B has won all its games against C, but C has beaten A 5 games to 4. Thus A has won 13 games, B has won 9, and C has won 5. From these results we must name a league champion. Most people would probably choose A, making B one of the "losers." But if B were suspended for hazing and became ineligible, and we therefore ignored the results of B's games, we would be forced to choose C, who had beaten A 5 games to 4. Thus the initial choice of A violates IIA. But is that reasonable? Team B's ineligibility doesn't mean that the results of its games are irrelevant to the choice between A and C.

Sports and elections are two different things, but Darlington (2017) reports a wide variety of simulation studies, all showing that voting systems satisfying IIA select worse candidates than systems violating IIA. Since the whole purpose of optimizing criteria is to promote selection of the best candidates, IIA seems counterproductive.

Darlington (2016, pp. 12-13) analyzed a minimax violation of the multiple-districts criterion. In that example, under minimax candidate A wins a 4-candidate race in each of two districts because within each district there is a large Condorcet cycle involving B, C, and D, similar to the cycle in the 605-voter example of Table 1 in Darlington (2016). But in the two districts the cycles are in opposite directions, so the cycles cancel each other out when the districts are merged, and a different candidate wins, thus violating the criterion. One possible defense of minimax here is that the example is artificial and contrived, and would rarely occur in real life. But a more fundamental point is as follows. The example in Table 1 of Darlington (2016) shows that there is sometimes a genuinely good reason to reject candidates involved in a strong Condorcet cycle. But in the two-district example the presence or absence of cycles depends on the area covered. Thus the example's most important lesson may be that with minimax it's more important than with other voting systems that votes be analyzed at the level of the total area served by the election's winner, not within subareas. That's normally done anyway; votes for the US president are one of the few exceptions. Therefore the multiple-districts criterion is not a good reason to reject minimax or other similar systems.

We have seen that there are good reasons to dismiss all the electoral criteria which conflict with Condorcet consistency. The next section shows the need for Condorcet consistency. As mentioned above, we start with majority rule.

## Why majority rule is essential

### *Majority Rule (MR), majority judgment (MJ), and range voting (RV)*

When we seek the best voting system for two-candidate elections, we can ignore the familiar impossibility theorems that bedevil the same search for multi-candidate elections. Thus limitations of any particular two-candidate system cannot be dismissed with the remark that every system has equally serious problems. This section shows that all two-candidate systems except MR have flaws serious enough to make them unacceptable, leaving MR as the only acceptable system for those elections. This conclusion goes well beyond a similar theorem proved by May (1952), since May's theorem doesn't cover systems like majority judgment (MJ) and range voting (RV) in which voters rate candidates on a multi-point scale, whereas such systems are covered here.

MJ and RV are the only well-known voting systems which don't reduce to MR for two-candidate races. They are also the only well-known systems in which each voter must rate each candidate on a scale, rather than ranking them or using some simpler system such as plurality or approval voting. RV uses a numeric scale, often from 0 to 99, and the candidate with the highest mean wins. MJ uses a scale with non-numeric verbal labels such as "excellent" or "poor," and the candidate with the highest median wins. If there is a tie in MJ, the ratings for each tied candidate are sorted from high to low. These sorted columns (one for each candidate tied for the win) can be assembled into a matrix. In that matrix we find the untied row nearest the median row, and the candidate with the highest rating in that row is the winner. If two such rows are equidistant from the median row, the one below the median row is used.

### *How an overwhelming favorite can lose under MJ or RV*

**Example 1** concerns MJ. It has two candidates A and B, 99 voters, and a 6-point scale with 1 low and 6 high. Suppose 49 voters rate A at 2 and B at 1, 49 others rate A at 6 and B at 5, and one voter rates A at 3 and B at 4. Then A's median rating is 3 and B's is 4, so B wins by MJ, although 98 of the 99 voters preferred A to B. In their book promoting MJ, Balinski and Laraki (2012, p. 328) mention this point, but dismiss it mostly because this situation has been demonstrated only in artificial examples which would rarely occur in practice. That defense may apply here, but not to the criticisms of MJ in later subsections. And the result is so bizarre that even a very occasional occurrence should be troubling. Situations like this would occur more often if some voters use primarily the top end of the scale while others use primarily the bottom end. And recall the earlier point that when focusing on two-candidate systems we can be less tolerant of anomalies than we might have to be in multi-candidate systems.

Balinski and Laraki (2012, pp. 328-329) claim that in two-candidate races, a candidate preferred by only one voter may also win in approval voting, in which each voter "approves" as many candidates as they wish, and the candidate with the most approvals wins. In the example they offer to prove their point (on their p. 281), 4 voters regard candidate X highly but regard Y even higher, and 4 others regard Y poorly but regard X even lower. A ninth "median" voter considers X slightly above average and Y and slightly below average. If all opinions above average are "approvals" and all others are "disapprovals," then X and Y get 5 and 4 approvals respectively. Thus X wins by approval voting, even though 8 of the 9 voters preferred Y to X.

This argument has two flaws. First, in real life a voter in a two-candidate approval election would realize they will have no effect if they approve both candidates or disapprove both, so they would approve only the one they regard more highly. Thus in real life, approval voting would reduce to MR in two-candidate elections. The second flaw in the argument is that the problem appears in approval voting only because of its extreme emphasis on simplicity. In this example X wins because the ballots don't even show that 8 of the 9 voters prefer Y, whereas in the earlier example MJ picks B even though the ballots show clearly that 98 of the 99 voters prefer A to B. Thus MJ but not approval voting is guilty of ignoring information that is right on the ballots. Using this argument to defend MJ, with its much more complex ballots, is like defending the presence of some fault in a luxury car on the ground that the fault is also found in one model of budget car.

**Example 2** concerns RV. It also has two candidates and 99 voters, but uses a rating scale from 0 to 99. Suppose 98 of the voters each rate A one point above B, but the 99th voter rates A at 0 and B at 99. Then B's rating total across voters exceeds A's by 1 point. Thus B wins by RV although 98 of the 99 voters preferred A to B. This example also illustrates the effectiveness of strategic voting in RV, since a single strategic voter who favors B would vote exactly as in this illustration, thus tipping an election in which every other voter prefers A.

### The effectiveness and safety of strategic insincere voting

The Gibbard-Satterthwaite theorem, discussed in many sources including Tideman (2006, pp. 143-148), asserts essentially that no reasonable voting system can always make it impossible to engage in successful strategic voting. But that theorem applies only to elections with 3+ candidates, it usually requires the insincere voter to know or guess accurately how others are voting, it often applies only in the presence of a Condorcet paradox (which is rare), and it is mostly concerned with mild forms of insincere voting such as a failure to rank all candidates even when a voter could do so. The flaws discussed here are much worse, and they arise even in two-candidate races.

**Example 3** involves voters who vote insincerely in two-candidate elections by giving their favored candidate the highest possible rating and giving the other candidate the lowest possible rating. That strategy will be obvious to most voters, can never backfire, and doesn't require voters to even guess how others are voting. Again we have 99 voters and two candidates. Voters are assumed to differ only on a single liberal-conservative opinion dimension we'll call LR for "left-right." Opinions on LR are normally distributed. Specifically, one artificial voter was placed at each of the 99 percentile points of a standard normal distribution. Candidate A was placed at 0 – exactly at the mean and median of the 99 voters. Thus A is the best possible candidate, because people in this model differ only on that one dimension. Candidate B was placed at +0.5. It worked out that 69 of the 99 voters fell below B, and the other 30 fell above. Thus B differed noticeably from the mean and median of the voters.

We'll consider three voting systems: RV, MJ, and MJD, which is a variant of MJ in which each voter's rating of a candidate is an exact linear function of the distance between the voter and the candidate. Thus the extra D in the name stands for "distance." The reason for using MJD is that, at least in this artificial example, ratings in MJD are rarely if ever tied with each other, so there is no need for a tie-breaker and candidates can be compared on their actual median ratings.

For RV and MJD, a voter's rating of a candidate was computed as 3 minus the distance between the voter and the candidate on scale LR. By this rule the leftmost 59 voters preferred A and the other 40 preferred B. The mean ratings of A and B were 2.224 and 2.125 respectively; their median ratings were 2.326 and 2.247 respectively. Thus A would be declared the winner by MR, RV, and MJD. The political news organization *Politico* has defined a "landslide" win as one with a 10% or higher victory margin by MR. The margin in this example is (59-40)/99 or 19.2%, so this qualifies almost as a "double landslide." The number of voters preferring A was 147.5% of the number preferring B.

The ratings just used were all sincere ratings. However, from the 40 voters who preferred B, suppose the 6 voters highest on LR (the ones with the lowest ratings of A) voted strategically by giving B the highest rating sincerely given to any candidate by any voter, and giving A the lowest rating sincerely given to any candidate by any voter. The other 93 voters voted sincerely as before. Now B's mean and median ratings both exceeded A's; the means for A and B were respectively 2.166 and 2.208, while the medians were 2.326 and 2.349. Thus in this example, six strategic voters out of 99 were enough to tip the election in B's favor by either RV or MJD, despite the landslide proportions of A's victory under sincere voting. If only the 5 voters highest on LR had voted strategically, MJD would have picked A while RV would still pick B. Thus in this example, RV is even more susceptible to strategic voting than MJD is, though both are unacceptably susceptible. Note that in all the examples of this section, all voters preferring B are already expressing that preference in their sincere votes, and we're considering just the difference between sincere and insincere expressions of that preference.

For those who want it, this paragraph gives more detail about MJD's behavior in this example. Of the 99 voters, the rightmost 40 were those who preferred B. The leftmost 15 *of those 40* gave both A and B sincere ratings above their respective medians. Those 15 could not help B by insincerely raising their ratings of B, because raising those ratings would not raise B's median. But they could help B by insincerely lowering their ratings of A, thus lowering A's median. The rightmost 15 of that same 40 (who were also the rightmost 15 of all 99 voters) gave both A and B sincere ratings *below* their respective medians. Those 15 could not help B by insincerely lowering their ratings of A, because lowering those ratings would not lower A's median. But they could help B by insincerely raising their ratings of B, thus raising B's median. The central 10 of those 40 voters (the 40 who preferred B) could not help B at all by voting insincerely. Those voters sincerely gave A ratings below A's median, and gave B ratings above B's median. If those voters were to rate A still lower, and rate B still higher, they would not change either of their medians. The bottom line is that of the 40 voters who sincerely preferred B, 30 could help B through insincere voting, either by raising B's median or lowering A's, though no single voter could have raised B's median and also lowered A's.

The MJ method usually employs far fewer scale points than were used above in RV and MJD. Therefore a separate MJ analysis was performed, using a scale with just 6 points. The ratings in the previous analysis ran from 0.174 to 3.000. Therefore, all values below 0.5 were changed to 1, those between 0.5 and 1.0 were changed to 2, those between 1.0 and 1.5 were changed to 3, and so on up at intervals of 0.5, so the highest transformed ratings were 6. The two candidates' medians were now tied, so MJ's tie-breaker was applied. The median row of the two-column matrix is row 50. With sincere ratings the closest untied row was row 39, in which A had a rating of 6 and B had 5, so A won. But when the process was repeated with ratings in which the 6 voters highest on LR had voted strategically to favor B, row 41 had a rating of 5 for A and 6 for B. That was now the untied row closest to row 50, so B now won. Thus 6 insincere votes out of 99 still tipped a landslide election.

As we have seen, examples 2 and 3 both suggest that RV is even more vulnerable to strategic voting than MJ. Two factors make RV more vulnerable than MJ. First, RV typically uses many more scale points than MJ, so insincerely rating one candidate at the very top, and the other at the very bottom, would have a larger effect in RV than in MJ. Second, we just saw that in a typical example, some of the voters preferring B cannot help B at all by insincere voting, and the rest can change one median in their preferred direction but not both. But in RV, every voter favoring B, except those whose sincere ratings are at the very top or bottom of the scale, can both raise B's mean and lower A's by insincere voting.

Will voters actually vote insincerely? Suppose a two-candidate election is run under MJ or RV, but for general interest or because some people demanded it, authorities also report the number of voters who rated each candidate above the other. Suppose voters then see that more voters favored the MJ or RV loser than the winner. That means that voters in the latter group had in some sense voted more effectively than the others. Those favoring the loser will immediately doubt the sincerity of the opposing voters, so in the next election they will almost certainly vote insincerely themselves. Soon everyone will be voting insincerely, giving their preferred candidate the highest possible rating and their opponent the lowest possible. Thus all two-candidate elections will reduce to MR.

### Insincere voting in other voting systems

It seems almost inevitable that results much like these would apply to any two-candidate voting system which isn't equivalent to MR. The essence of MR is that all voters favoring a particular candidate are treated equally; those who love candidate A and loathe B help A no more than those who prefer A just slightly over B. Thus any reasonable system not equivalent to MR must somehow treat two such voters differently, presumably giving more weight to the former. But when voters know that some response patterns will be given more weight than others preferring the same candidate, they will have every incentive to insincerely give the response pattern with the higher weight.

It seems clear that any system comparing measures of central tendency, as MJ and RV do, will be at least as vulnerable to insincere voting as MJ is. Measures of central tendency differ in the amount they are influenced by extreme scores. For instance, a group's midrange is the mean of the highest and lowest scores in the group; it's obviously heavily influenced by extreme scores. Of all measures of central tendency, the median is the one least influenced by extreme scores. If we change any score other than the original median score, while keeping the changed score on the same side of the median, it will not change the median at all. Thus it seems likely that MJ will be less vulnerable to insincere voting than systems which use almost any other measure of central tendency. But we have already seen that MJ itself is unacceptably vulnerable to insincere voting.

### Majority rule estimates more efficiently than MJ

RV is more susceptible than MJ to insincere voting because RV uses data which MJ ignores. But that difference also means that MJ is surprisingly inefficient when there are no insincere voters or other irregularities. This section illustrates that point.

This analysis was like the previous analysis in several ways. Again, voters in a population were assumed to have a standard normal distribution on the LR scale used above. Candidate A's position on LR was again at 0, which is exactly the mean and median of all voters. People in this model again differ only on that one dimension, so A is still the ideal candidate. In this study, candidate B was placed at different positions on LR in different parts of the study; B's position was 0.1 or 0.2 or 0.3 or 0.4 or 0.5. In each trial a random sample of voters was drawn from the population. Each voter's rating of each candidate was again assumed to equal 3 minus the absolute scale difference between the voter and the candidate. These ratings were then analyzed by both MR and the MJD system described earlier. Under MJD the candidate with the higher median rating was the winner. Under MR, each voter voted for the candidate nearer him or her on the opinion dimension, and the candidate with the most votes won.

The study had 3 sections, each with 5 subsections. There were 10,000 mutually independent trials in each subsection. In sections 1, 2, and 3, the numbers of voters were respectively 15, 55, and 95. In subsection 1 of each section, candidate B was at 0.1 on the scale. In subsections 2-5, B was respectively at 0.2, 0.3, 0.4, and 0.5. As already mentioned, A was always at 0. Since A was in fact the ideal candidate in all trials, the measure of a system's excellence was the number of trials on which the system picked A as the winner. The only trials reported here were trials in which one system picked A while the other system picked B. MR outperformed MJD in all 15 subsections of the study. The larger the number of voters, and the farther B was from the mean of 0, the wider MR's margin over MJD. When there were 95 voters and B was at 0.5, there were 1532 trials in which MR picked A while MJD picked B, and only 94 trials in which the reverse was true. Thus MR was "right" over 16 times as often as MJD when the two methods picked different winners. The smallest difference between the methods occurred with the smallest number of voters (15) and the lowest position of B (0.1). Then MR outperformed MJD on 2106 trials and the opposite occurred on 1716 trials; the former number is 23% larger than the latter. Thus MR always noticeably outperformed MJD, and the difference between the two systems increased with the number of voters and the difference between the candidates.

# References

Balinksi M, Laraki R (2010) Majority judgment: measuring, ranking, and electing. MIT Press,  Cambridge MA

Darlington RB (2016). Minimax is the best electoral system after all. https://arxiv.org/ftp/arxiv/papers/1606/1606.04371.pdf

Darlington RB (2017). In elections, irrelevant alternatives provide relevant data. http://arxiv.org/abs/1706.01083

Felsenthal DS (2012) Review of paradoxes afflicting procedures for electing a single candidate. In: Felsenthal DS, Machover M (eds) Voting systems. Springer Verlag, New York

Laslier J-F (2012) And the loser is…plurality voting. In: Felsenthal DS, Machover M (eds) Voting systems. Springer Verlag, New York

May, Kenneth O. 1952. A set of independent necessary and sufficient conditions for simple majority decisions, *Econometrica*, Vol. 20, Issue 4, pp. 680–684.

Tideman TN (2006) Collective decisions and voting: the potential for public choice. Ashgate, Burlington VT